\documentclass[twocolumn,aps,pra,floatfix]{revtex4}

\usepackage{graphicx,graphics}
\usepackage{times}
\usepackage{psfrag}

\newcommand{\be}{\begin{equation}}
\newcommand{\ee}{\end{equation}}
\begin{document}

\preprint{Submitted to Phys. Rev. B}

\title{
Mode transformations and entanglement relativity in bipartite gaussian states
}


\author{Emanuele Ciancio $^1$}
\email[]{ciancio@isiosf.isi.it}
%\thanks{}
\author{Paolo Giorda$^{1,2}$}
\email[]{giorda@isiosf.isi.it}
%\thanks{}
\author{Paolo Zanardi$^1$}
\email[]{zanardi@isiosf.isi.it}
%\thanks{}
\affiliation{$^1$
Institute for Scientific Interchange (ISI),
Viale Settimio Severo 65 ,
10133 Torino, Italy
}
\affiliation{$^2$ Dipartimento di Fisica del Politecnico, C.so
Duca degli Abruzzi 24, I-10129 Torino, Italy}


\date{\today}

\begin{abstract}
A proper choice of subsystems  for a system of identical particles
e.g., bosons, is provided by second-quantized modes
i.e.,creation/annihilation operators. Here we  investigate how the
entanglement properties of bipartite gaussian states of bosons
change when modes are changed by means of unitary, number
conserving, Bogolioubov transformations. This set of "virtual"
bi-partitions is then finite-dimensionally parametrized and one  can
quantitatively address relevant questions such as the determination
of the minimal and  maximal available entanglement. In particular,
we show that in the class of bipartite gaussian states there are
states which remain separable for every possible modes redefinition,
while do not exist states which remain entangled for every possible
modes redefinition
%(absolutely
%entangled).
% ***
%absolutely separable states do exist whereas absolutely entangled do not.
\end{abstract}

% insert suggested PACS numbers in braces on next linemanual
%\keywords{}

\maketitle

%\section{}
%\subsection{}
%\subsubsection{}

% If in two-column mode, this environment will change to single-column
% format so that long equations can be displayed. Use
%\begin{widetext}
%\end{widetext}

\section{Introduction}\label{s:intro}
%INTRODUCTION

%\documentclass{article}
%
%\begin{document}
%
% {}
%
%\title{PRELIMINAR RESULTS}
%\maketitle

%\section{Introduction}
The physical notion of quantum entanglement lies at the very heart
of the novel and rapidly developing discipline of quantum
information science \cite{qis}. The  mathematical definition of
entanglement requires the existence of a tensor product structure
over the Hilbert state-space of the system $S$ under investigation;
physically a multi-partition of $S$ into subsystems has to be
identified. It is nearly obvious that this latter choice is by far
not unique; in most of the cases  there is actually a enormous set
of potential multi-partitions available. Unless a physical, possibly
problem dependent selection criterion is somehow provided, all those
"virtual" multi-partitions of $S$ into subsystems are conceptually
on the same foot.

In Refs \cite{virtual} and~\cite{QTP} an operationally motivated
framework for relativity of entanglement, based on operator-algebras
of observables has been introduced. There, the key physical idea is %***
that the relevant multi-partition is selected on the basis of the
set of implementable quantum dynamics, including both unitaries and
measurements, in the actual operational scenario. Mathematically,
the necessary tensor product structure is {\em induced} over the
global set space by a bunch of implementable commuting algebras of
observables satisfying  suitable natural properties \cite{QTP}. Each
of this algebras becomes then a {\em local } one i.e., it plays the
role of the observable algebras of one of the "virtual" subsystems.

A very important class of physical systems is provided by the
ensembles  of identical particles. For this kind of systems, in view
of the (anti)symmetrization postulates used to enforce the proper
statistical behaviour, a  tensor product structure of the physical
state-space e.g., the totally symmetric subspace for bosons, is not
naturally given. Accordingly, the extension of the notion of quantum
entanglement to systems of indistinguishable particle is not a
totally trivial task and it indeed gave raise to different,
complimentary approaches \cite{ferm}, \cite{part}. In this paper we
shall follow the second-quantized approach where the subsystems are
associated  to {\em modes} \cite{ferm}, rather than to particles
\cite{part}. We will focus on bi-partite bosonic gaussian states and
we will study how entanglement changes under unitary,
number-conserving, redefinition of modes i.e., subsystems. Gaussian
states play a central role in the theory of entanglement for
continuous variables
systems; % of continuos variables; ***
the reason is two-fold. First their experimental relevance; indeed
most of % ***
the states of light, which are easily obtainable in the lab by means of linear optics
devices e.g., beam splitters, are gaussian. Second, they are
mathematically easy to handle. In spite of the fact that gaussian
states live in the infinite-dimensional bosonic Fock space all of
their physical features are indeed captured by  the so-called {\em
Covariance Matrix} $V$. From %***
this {\em finite-dimensional} matrix one is able to extract
straightforwardly various measures of entanglement like negativity
and logarithmic negativity \cite{VW}. To this aim it is often useful
to cast the covariance matrix in standard form  $V_s.$ This is
possible via local operations
which do not change the entanglement of the state \cite{StanFor}.

The main goal of our work is to focus on the idea of {\em
entanglement relativity} \cite{virtual,QTP} as it can be applied in
the field of gaussian states. As already pointed out, once a
gaussian state is given in terms of its covariance matrix with
respect to a given tensor product structure, we study how the
entanglement properties of {\em this} state  change when new tensor
product structures are defined via number conserving modes
transformations. The latter will be defined in detail in the
following sections. It is important to emphasize that this paper is
not primarily aimed at studying gaussian states {\em per se}. The
latter are used, thanks to their structural simplicity, as an ideal
arena to address and illustrate in a quantitative way the somewhat
abstract idea of entanglement relativity.   It is important to
stress the difference, with respect to the existing literature, of
the conceptual status of the modes transformations that we are going
to
consider. Indeed, %***
in most of the former studies (e.g., \cite{WEP}), modes
transformations are physically enacted by optical devices (e.g.,
beam-splitters) that actually {\em change} the quantum state under
examination. In this paper instead, the quantum state $\rho$ of two
bosonic modes, has to be thought as {\em unchanged}: it is just the
way $\rho$ is decomposed into subsystems i.e., modes, that is
modified. The situation is somewhat reminiscent of the difference
between "active" and "passive" rotations in elementary geometry;
just as in that case, while the mathematical objects involved
can be the same, the conceptual interpretation is quite different.%##

The paper is organized as follows. In Sect. (\ref{sec:er}) the basic
ideas of entanglement relativity are reviewed. In Sect.
(\ref{sec:gs}) theoretical background on gaussian states is
provided. Sects (\ref{sec:ps}) and (\ref{sec:ms}) are devoted to the
analysis of pure and mixed gaussian states respectively. Finally
Sect. (\ref{sec:conclu}) contains some discussions and conclusions.

\section{Entanglement relativity}\label{sec:er}
In the general setting \cite{QTP} the local commuting algebras we are going to consider in the following are those
generated by the mode creation/annihilation operators \cite{ferm}; the global state-space considered here is the Fock space
(total number of particle unconstrained) that  it is known to be isomorphic to the tensor product of harmonic oscillator
%#
state spaces.
It is an elementary fact that, by  using canonical transformations, one can move from a given set of modes
to a new one defining a new tensor product structure i.e., new subsystems.
One is then  naturally led to analyze how the entanglement properties of a given quantum state $\rho$
depend on the  choice of the set of modes and how they change when a different set of modes is considered.
%#
In particular, once $\rho$ is given, one may want to know whether it
exists a choice of modes i.e., a tensor product structure, with
respect to which $\rho$ is a separable state. In the negative case
%#
one may refer to such a state as {\em absolutely} entangled.
%#
Another natural task is to find  the minimum and the maximum of
the entanglement of $\rho$ over the set of all possible tensor product structures
(i.e, modes redefinitions). Besides its theoretical interest this might have some practical
relevance. For instance,the latter analysis
allows one to identify for any given state the
modes that correspond to the maximal entanglement and that can be
fruitfully exploited for some quantum information processing protocol e.g.,
teleportation.

More formally, if one denotes by $\cal T \in {\cal TPS}$ one of all the possible tensor product structures (TPS) %over $S$
and by $E_{\cal T}$ an entanglement measure e.g., negativity, associated to the TPS $\cal T$ one can define
at least two natural "absolute" i.e., TPS-independent, entanglement measures
\begin{eqnarray}
E^{+}(\rho)&:=& \sup_{{\cal T}\in{\cal TPS}} E_{\cal T}(\rho),\nonumber\\
E^{-}(\rho)&:=& \inf_{{\cal T}\in{\cal TPS}} E_{\cal T}(\rho).
\label{abs-meas}
\end{eqnarray}
The set of absolutely separable \cite{Zyc} and absolutely entangled
states would then correspond respectively to $\{\rho\,/\,
E^+(\rho)=0\}$ and $\{\rho\,/\, E^-(\rho)>0\}.$ As mentioned above a
very natural question that arises concerns the relative weight of
these two subsets in full state space. Of course, a more pratically
meaningful definition of a TPS-independent entanglement measure
would involve in Eq. (\ref{abs-meas}) a restricted set of
operationally meaningful TPSs. Indeed, it should be clear that the
definitions (\ref{abs-meas}) will in general involve optimization
procedures over a vast set of possibilities.

In order to make this
program amenable to a quantitative treatment, in this paper we shall
consider
bi-partite gaussian states only. This allows us to focus
on a subset of ${\cal TPS}$ whose elements are parametrized by
finite-dimensional objects i.e., $N$ by $N$ unitary matrices.

\section{Theoretical background}
\label{sec:gs}
\subsection{Gaussian states}
Gaussian states are defined as those
quantum states whose
Wigner function is gaussian;
they are completely defined by their second
statistical moments.
By grouping all the components $x_i$
of the position and momentum operators in a single column vector such that
$x_i=q_i,x_{i+1}=p_i$,
one can write \cite{Sim}:
\begin{equation}\label{Wig}
W(x) = (4\pi^2 \sqrt{detV})^{-1} exp \left[ -{1 \over 2}(x^{t}
V^{-1} x) \right]
\end{equation}
where $W(x)$ is the Wigner function and $V$ is the covariance
matrix. The entries of the latter are exactly all the second order
statistical moments of the components of the canonical operators
(position and momentum).

In general, in terms of the components $x_i$, the classical
covariance is defined as: $< x_i x_j > - <x_i><x_j>$. Dealing with
quantum operators the products appearing in the covariance
expression must be taken in the Wigner-Moyal form, i.e. they must be
symmetrized: $ < x_i x_j >  = < x_i x_j + x_j x_i > /2$.
Since we work
with gaussian states, the first moments (mean values) can be set to
zero
by means of displacement operators
that do not change the
entanglement properties \cite{GPS}.
Therefore $V$ is built by simply
setting: $V_{ij} = < x_i x_j >$. In this way
, the covariance matrix
$V$ has all the variances $< x_i^2 >$ on the diagonal and the covariances
$< x_i x_j >, \ i\neq j$ off diagonal.
Due to the Wigner-Moyal
symmetrization
$V$ turns out to be real and symmetric.
In the following we will deal with
bipartite systems: the identification of $x_1,x_2$ with $q_A,p_A$ and
of $x_3,x_4$ with $q_B,p_B$, allows us to write $V$
in a block form:
\begin{equation}
V =
\left(
\matrix{A & C \cr
C^t & B}
\right)
\label{covmat}
\end{equation}
where $A,B$ refers  to subsystem A(lice) and B(ob) respectively.

By using the covariance matrix formalism the  %***
Heisenberg uncertainty principle can be recast in the following compact way:
\begin{equation}
V \ge {i \over 2}\sigma,
\end{equation}
where $\sigma$ is the so-called symplectic unity for two subsystems:
\begin{equation}
\sigma = \left( \matrix{ 0 & 1 & 0 & 0 \cr -1 & 0 & 0 & 0 \cr 0 & 0
& 0 & 1 \cr 0 & 0 & -1 & 0 } \right) \label{sigmatrix}
\end{equation}
Here, the uncertainty principle is expressed in its extended form;
together with the variance relations, also the covariance relations are
taken into account: $Var[x_i] Var[x_j] - Cov[x_i,x_j]^2 \ge 1/4$.

Any covariance matrix $V$ can be put in the so called {\it standard form} $V_s$
by means of a suitable {\it local symplectic transformation} (see below).
In $V_s$ the blocks have a diagonal form \cite{StanFor}:
\begin{equation}
V_s =
\left(
\matrix{a & 0& c_+ & 0 \cr
0 & a & 0& c_-\cr
c_+ & 0 & b & 0\cr
0 & c_-&0 & b}
\right).
\label{Vstand}
\end{equation}
Since the transformation that gives $V_s$ is local, it does not change the
entanglement properties of the state.

\subsection{Modes transformations}

A generic bosonic bipartite (gaussian) state can be described in
terms of two harmonic oscillator modes. In a quantum system the
operators accounting for the oscillation amplitudes are the
annihilation/creation operators $a_i, a^{\dagger}_i$. The two
subsystems corresponding to the bipartition are the two oscillation
'directions'. We can imagine Alice and Bob sharing the system, Alice
having one mode, Bob having the other one. According to what we have
already said about the relativity of the entanglement, one state can
be entangled if described by a particular pair of modes and can be
separable if described by a different pair. If we now consider how
the modes can be changed we see that, from a phase-space point of
view, the modes transformations correspond to canonical
transformations of the conjugated  (position and momentum)
hamiltonian variables. The new variables must of course fulfill the
same canonical commutation relations. The set of these canonical
transformations form the symplectic non-compact group:
$Sp(2n,\cal{R})$.

A well known class of mode transformations are the so called Bogolioubov transformations:
$a \rightarrow b, \ \ b_i = \sum_j \alpha_{ij} a_j + \beta_{ij} a^{\dagger}_j$, with $i,j = 1,...,n$.
For bipartite states $n=2$.
A special sub-class of Bogolioubov transformations is given by:
\begin{equation}
 b_i = \sum_j U_{ij}a_j, \quad (i,j=1,2)
\label{mode:redef}
\end{equation}
where $U$ is a unitary $2\times 2$ complex matrix. This kind of mode
transformations are those which conserve the number of particles and
are precisely the ones we are going to focus on
in our analysis.
%Given a particular state, a mode transformation can change the entanglement of the state.
% (i.e. its negativity)
%An additional
%question is then addressed: 'Are there states which remain entangled/separable
% ($\cal{N} \ne 0$ / $\cal{N} = 0$)
%for every possible mode transformation?'.
These mappings form a group  $K(n)$  that coincides with the compact
subgroup of $Sp(2n,\cal{R})$ of maximal dimension \cite{symple}.
Each mapping (\ref{mode:redef}) induces a specific tensor product
structure  $\cal T \in {\cal TPS}$. The goal of our paper is then to
study  how  entanglement varies when the modes redefinitions
belonging to
$K(n)$ are applied to a given state. %^
In particular
this means that in the remainder of the paper the ensemble  ${\cal
TPS}$ used to evaluate (\ref{abs-meas}) is in a one-to-one
correspondence with elements of $K(n)$.

Each of the unitary and symplectic transformations $S_U \in K(n)$, where $U$ is  defined in
(\ref{mode:redef}), can be written as ~\cite{nota}:
\begin{equation}
S_U =
\left(
\matrix{X & -Y \cr
Y & X}
\right)
\label{uniS}
\end{equation}
where $X$ and $Y$ are respectively the real and the imaginary part
of the unitary $2 \times 2$ operator $U=X+iY$ acting on the  modes
\cite{symple}:
\begin{eqnarray}
X = {U+U^* \over 2} \nonumber \\
Y = {U-U^* \over 2i}.
\end{eqnarray}
The covariance matrix transforms according to (\ref{uniS}) as
\be
V'=S_UVS_U^t. \label{Vtrans}
\ee
We end this subsection by
discussing how one can represent a general element of $K(n)$. In
general a  $2 \times 2$ unitary matrix $U$ depends on four real
parameters; thus a possible and useful parametrization of $U$ is the
following:
\begin{equation}
U =
\left(
\matrix{
\sin\theta e^{i\phi} & \cos\theta e^{i\phi'} \cr
\cos\theta e^{i\phi''} & \sin\theta e^{i(-\phi+\phi'+\phi''-\pi)}
}
\right).
\label{Unipars}
\end{equation}
The latter is the parametrization we use in our numerical calculations.

\subsection{Entanglement measure}
%The entanglement properties of quantum states have been deeply studied in the
%past years. The entanglement has been indicated as the key feature that ultimately distinguish
%a quantum system from a classical one. It is also has been recognized to be at the heart of
%many important quantum communication protocols (refs???).
%It is then important to properly
%describe it.
In continuous variables systems a proper measure of entanglement
that can be efficiently computed is logarithmic negativity
\cite{Peres,HHH,VW}. Following Ref.~\cite{VW}  logarithmic
negativity can be obtained starting from the trace-norm of the
partially transposed density matrix. This trace-norm  can be
computed from the so-called symplectic eigenvalues of the partial
transpose of $V$ with respect to the subsystem $A$, i.e. the
eigenvalues of $\sigma^{-1}V^{T_A}.$ The characteristic equation for
$\sigma^{-1}V^{T_A}$ is:
\begin{equation}
\lambda^4 - (\det A +\det B - 2 \det C)\lambda^2 + \det V = 0
\end{equation}
Thus the logarithic negativity  can be computed as:
\begin{equation}
E_{\cal{N}} = {1 \over 2} \sum_i \max[-\log_2(2|\lambda_i|),0].
\label{neg}
\end{equation}

\section{Results}

In order to give a quantitative description of how the entanglement
of a gaussian state changes under the action of all unitary  transformations
$S_U \in K(n)$,
we have performed different kinds of numerical computations that now we briefly describe.

{\em I. Pure states} We start our analysis by considering pure
gaussian states whose covariance matrix is in the standard form
$V_s$. % The latter depends only on one real parameter $a \ge 1/2$.
The modes transformations are generated by using a suitable
discretization of the four dimensional parameters space defining
$U$, see (\ref{Unipars}). For each transformation (mode
decomposition) we compute the logarithmic negativity of the
corresponding covariance matrix (\ref{Vtrans}).

\begin{figure}[h]
\psfrag{S}{\small $S_U$}
\psfrag{V}{\small $V_s$}
\psfrag{T}{\small $T$}
\psfrag{R}{\small $\tilde{V}$}
\psfrag{F}{\small Fixed entanglement}
\psfrag{E}{\small Fixed energy}
\fbox{\includegraphics[height=5cm, width=8cm, viewport= 30 190 700
710, clip]{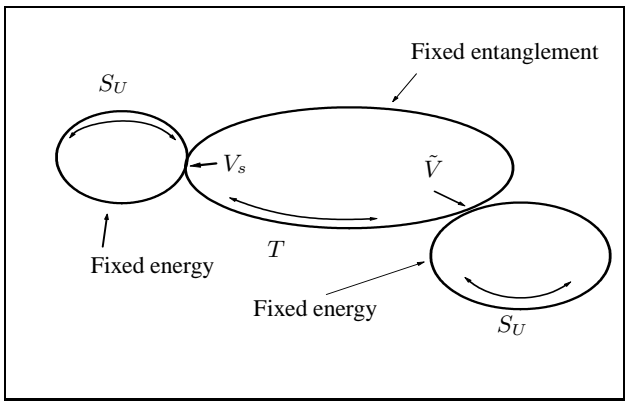}}
\caption{
A visual display of the procedure used to generate and decompose the gaussian states.
}
 \label{Anelli}
\end{figure}

Obviously the covariance matrices in the standard form do not
describe all the possible gaussian states. $V_s$ is useful because
it depends only on four parameters and has the same entanglement of
a large class of states. % ***
Given a generic gaussian state and its covariance matrix $V$ it is
easy to bring the latter in standard form by applying a suitable
local symplectic transformation \cite{StanFor} $L \in Sp(2,
\mathcal{R})\oplus Sp(2,\mathcal{R})$ such that
$V_s = LVL^t$.
One can devise
the following
procedure to generate a set of generic $V$s starting from the $V_s$:
given a covariance matrix $V_s$ one can apply the
full set of local symplectic transformations $T \in Sp(2,
\mathcal{R})\oplus Sp(2,\mathcal{R})$; for each
$T$ one obtains the matrix $\tilde{V} = TV_sT^t$ that corresponds to
a squeezed state with the same entanglement of $V_s$ and with a possibly
different energy.
Once a generic $\tilde{V}$ has been
generated, we apply the modes redefinition scheme $S_U\tilde{V}S_U^t$ and we
evaluate the corresponding logarithmic negativity. The procedure is then repeated
for many different $V_s$ and it is schematically
displayed in Fig. (\ref{Anelli}).

We would like to underline that while the $S_U$s transformations
(also referred in the literature as {\em passive}) are energy
conserving, the $T$s are entanglement conserving and that in general
$S_UT \ne TS_U$. It is thus important  to stress that applying the
mode decomposition analysis to covariance matrices in non-standard
form allows us to access a wider class of results. In this respect
the present work differs from what has been done in several studies
on gaussian state entanglement, e.g. \cite{WEP,illu}. Those studies
were indeed aimed at analyzing just entanglement properties of
gaussian states, therefore the restriction to $V_s$ did not prevent
the achievement of  full generality.
Here our aim is to study the
variation of entanglement for different bipartitions associated with
unitary modes redefinitions, no matter if the initial covariance
matrix $V$ is $V_s$ or $\tilde{V}$. Thus the $T$s transformations
are merely used as a tool to define a more general set of new
gaussian states; accordingly they do {\em not} have to be regarded
as a new set of (number non-conserving) Bogolioubov transformations
applied to the initial $V_s$.``

{\em II. Mixed states.} A great number of mixed states have been
generated in a random way by drawing the parameters $a,b$ and
$c_\pm$ out of a uniform distribution over a well defined parameters
range (see caption Fig. \ref{stat1}). Each (real symmetric) matrix
is checked in order to verify whether it fulfills Heisenberg
uncertainty principle; this allows one to select only the physical
covariance matrices. The mode redefinition procedure is then applied
and the maximal and minimal entanglement achievable is computed, see
Eq. (\ref{abs-meas}). These computations have involved covariance
matrices both in standard and non-standard form.

In the following sections we describe in details the results of our computations.

\subsection{Pure states}
\label{sec:ps}

The first set of gaussian states we take into account%# are the pure states
whose covariance matrix
is in the standard form $V_s$.
In terms of $V_s$ they are described  by only one independent parameter.
In fact, with the usual notation (\ref{Vstand}), pure states correspond to
 covariance matrices  with $a=b$ (pure states are always symmetric),
 $c_{\pm} = \pm \sqrt{a^2 - 1}$, with the constraint  $a \ge 1/2$.
The pure gaussian states fulfill the Heisenberg uncertainty
principle with the equal sign
i.e., they minimize the uncertainty
relations.
We can consider two cases. In the first one  $c=0$: all covariance terms
are zero and the
corresponding states minimize the uncertainty relations in the usual
form ($\Delta q_i \Delta p_i = 1/2 $). In the
second case, $c \neq 0$, the pure gaussian states
corresponding to $V_s$ are the squeezed states \cite{illu}. It's not
difficult to see that the non-squeezed states ($c = 0$) correspond
to the isotropic oscillator case.

\noindent The results of the numerical computations  show that :
\begin{itemize}

\item[$i)$]
there is always a unitary redefinition of the modes
for which any pure state is separable i.e., $E^{-}(\rho)=0$;
\item[$ii)$]
in particular, the non-squeezed  states are {\em absolutely separable}: $E^{+}(\rho)=0$;
\item[$iii)$] in the case of squeezed states all unitary transformations
lead to redefinitions of the modes ({\cal TPS}) for which
the value of the entanglement is always
lower than value corresponding to the initial $V_s$.
\end{itemize}

\begin{figure}[h]
%\{
\includegraphics[height=5cm, width=8cm, %viewport= 10 230 700 780,
clip]{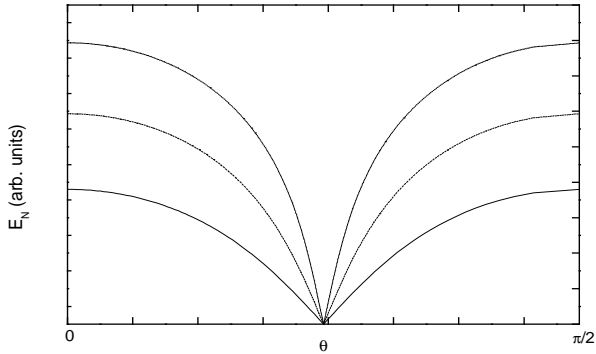}
%}
\caption{ Logarithmic negativity as function of the parameter
$\theta$ for different pure states. The picture clearly shows a
symmetric periodic behaviour. The point of zero entanglement is
reached for $\theta = \pi/4$ while the maxima are in correspondence
of $\theta = 0$, $\theta=\pi/2$. }
 \label{fips1}
\end{figure}

A few comments are in order. The existence of a unitary redefinition of the modes for
which the state is separable, see $i$), is not surprising. In fact
this can be understood, by elementary arguments, in different ways.
One one hand one can resort to the situation in which  two coupled
oscillators can be decoupled by the classical normal modes
decomposition. A suitable choice of the mode operators (either
classical or quantum) always allows one to deal with two independent
mode of oscillation.
%Given a generic
%\begin{equation}
%H = \hbar \omega_{11} a_1^{\dagger} a_1 + \hbar \omega_{22} a_2^{\dagger} a_2 +
%\hbar \omega_{12} a_1^{\dagger} a_2 +  \hbar \omega_{21} a_2^{\dagger} a_1
%\end{equation}
%it is always possible to find a transformation $a_i = \alpha_i b_1 + \beta_i b_2$ such
%that
%\begin{equation}
%H \rightarrow H' = \hbar \Omega_1 b_1^{\dagger} b_1 + \hbar \Omega_2 b_2^{\dagger} b_2
%\end{equation}
%where $\Omega_i = \alpha^2_i \omega_{11} + \beta^2_i \omega_{22}$.
%In the particular case in which $\omega_{12} = \omega_{21} = 0$ and $\omega_{11} = \omega_{22}$
%(isotropic oscillator):
%\begin{equation}
%H' = \hbar \omega_{11} b_1^{\dagger} b_1 + \hbar \omega_{22} b_2^{\dagger} b_2
%\end{equation}
%for every unitary transformation $U$.
%This means that we can consider
%the non-squeezed  states are absolutely separable;
On the other hand, the state is described  in the configuration
space by the Wigner function $W$ of Eq.~(\ref{Wig}). If we consider
$W$ as a function of $q_A, q_B$ and a plane parallel to the $q_A,
q_B$ plane intersecting $W(q_A, q_B)$,
%for different pure sates $\rho$
we get an ellipse whose axes lay on the bisectrices of the
quadrants. In the isotropic case ($|c_{\pm}| = 0$), see $ii)$, the
ellipse reduces to a circle, therefore the modes redefinitions, correspondig to rotations in the plane,
 do not
change the variances of $q_A, q_B$. When $|c_{\pm}| > 0$ a
covariance term appears
and it is responsible for a non-vanishing
excentricity of the ellipse, i.e. the gaussian Wigner function is
'squeezed'.
A rotation  (mode redefinition) of the reference axes
always allows one to make them conincide with the ellipse's ones; therefore, it is always possible
to set to zero the covariance term, which is responsible for the entanglement
between the modes. In our case a $\pi/4$
rotation reduces the entanglement to zero (see Fig. \ref{fips1}). In
the $q_{A(B)} p_{A(B)}$ plane instead, since $Cov[q_{A(B)},p_{A(B)}] = 0$ ( i.e., the blocks $A$ and $B$
in Eq. (\ref{covmat}) are diagonal), the  argument of
Eq.~(\ref{Wig}) simply identifies a circle.

We now turn to examine point $iii)$. In Fig. \ref{fips1}
the logarithmic negativity (\ref{neg}) of different pure
sates $\rho$ is plotted as a function of  $\theta$
($\phi,\phi',\phi''$ in Eq. (\ref{Unipars}) are kept fixed).

The maximal value of $E_{\cal{N}}$ i.e., $E^{+}(\rho)$,  is achieved in correspondence
of $\theta = 0$ (modes swapping) and $\theta=\pi/2$ (identity transformation).
Any other value of $\theta$ produces a lower value of $E_{\cal{N}}$.
In particular, for every state $\rho$, there is a value of $\theta$ for which $E_{\cal{N}} = 0$.
The results, that would not qualitatively change with a different choice of $\phi,\phi',\phi''$,
can be understood
as follows.
In case of symmetric states the maximal entanglement is achieved in correspondence of $V_s$
because, in this case,
 all the quantum correlations between the operators are set to zero except those involving same-type
operators between different subsystems. Every unitary modes
redefinition gives
rise to correlations between conjugated operators
belonging to the same subsystem, while it decreases the correlation
terms and consequently the entanglement between the two subsystems .
Similar results were obtained in the context of optical passive
transformations on gaussian states by Wolf et al.\cite{WEP} and they
are in agreement with the fact that the state described by $V_s$
corresponds, at given energy, to the maximum entanglement
\cite{Paris}.

We now turn to discuss the states whose covariance matrix $\tilde{V} = TV_sT^t$ is not in standard form.
It is important to realize that, at variance with the former studies of gaussian state entanglement,
in the present setting the passage to the standard form is {\em not} harmless. In fact in order to cast
a generic covariance matrix $V$ into its standard form $V_s,$ one has to perform local operations $L$
( e.g., local squeezing)
that are {\em outside}
the class of the  modes redefinitions here allowed
(\ref{uniS}).
Since these latter  do not in general commute with the local operations $L$ required to get $V_s$
one obtains, as we will now see, rather different results
depending whether the standard form is considered or not.

\begin{figure}[h]
%\fbox{
\includegraphics[height=5cm, width=8cm, clip]{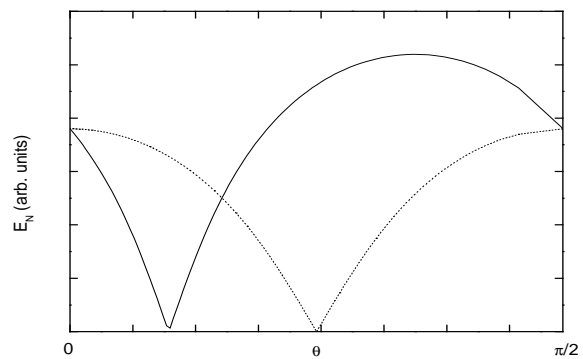}
%}
\caption{ The main difference between the behaviour of the pure
state with  $V_s$ (dotted line) and the one with $\tilde{V}$ (solid
line) is the possibility for the latter to increase its logarithmic
negativity }
 \label{pns}
\end{figure}

\begin{figure}[h]
%\fbox{
\includegraphics[height=6cm, width=6cm, clip]{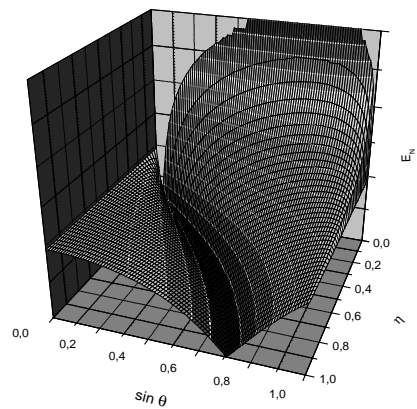}
%}
\caption{ A 3-dimensional plot of the logarithmic negativity as
function of $sin \theta$ and $\eta$. It is possible to note a line
laying on the ground plane (dark valley) testifying the
zero-entanglement modes redefinition for each initial state. Also it
is worth noting the increasing value of $E_{\cal{N}}$ as $\eta$ gets
different from $1$, for some $U \neq \openone$. }
 \label{3d}
\end{figure}

The numerical computations, see Fig. \ref{pns}, show that:
\begin{itemize}
\item[$iv)$]
as described for the previous case, it is always possible to find a bipartition in which the state is separable
 i.e., $E^{-}(\rho)=0$;
\item[$v)$] it always exists a redefinition of the modes that can increase
$E_{\cal{N}}$.
\end{itemize}
Fig. (\ref{3d}) shows the behaviour of the logarithmic negativity as a function of one of the local symplectic
parameter (the squeezing operation) $\eta: a \openone \rightarrow diag(\eta a,\eta^{-1} a,\eta a,\eta^{-1} a)$
and of one of the global unitary transformation parameter $\theta$.

\subsection{Mixed states}
\label{sec:ms} 
\begin{figure}[h]
\includegraphics[height=5cm, width=8cm, clip]{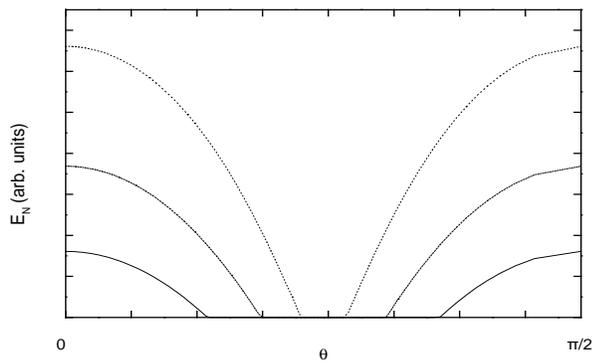}
\caption{
The plot of the logarithmic negativity as a function of $\theta$ for three different symmetric mixed states
shows a qualitatively similar behaviour for the three states. The width of the vanishing entanglement
plateau is related
to difference between $|c_{\pm}|$ and $|c^{ps}_{\pm}|$
}
 \label{ms}
\end{figure}

\begin{figure}[h]
\includegraphics[height=5cm, width=8cm, clip]{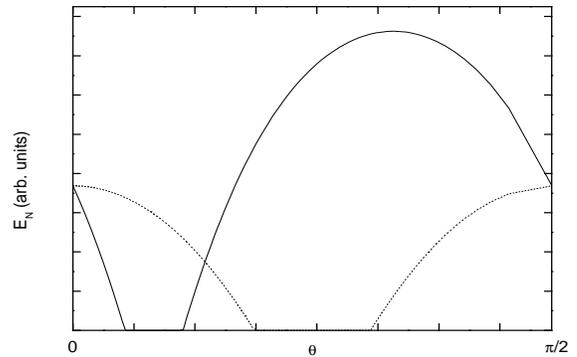}
\caption{ Logarithmic negativity as a function of $\theta$ for a
mixed symmetric state (solid line) with covariance matrix in
non-standard form, compared to the one of the corresponding state
with $V_s$ (dotted line).}
 \label{mns}
\end{figure}

The numerical analysis of mixed states  leads to
other interesting results.
As in the previous section we start by considering the
the covariance matrices in standard form. We  first look at the symmetric ($a=b$)
subset of mixed states depending only on three parameters. If we
set, as an additional condition, $c_+ = - c_-$, we have states that
behave in a way similar to the pure ones. In fact $E_{\cal{N}}$ is
maximal in correspondence of the identity transformation and it
decreases for all entangling  $U \neq \openone$. The difference
consists in the fact that, instead of a single point, there is a
whole interval of values of $\theta$ for which the logarithmic
negativity vanishes, as displayed in Fig. \ref{ms}.
The further one
moves away from pure states ($ps$)
 ($|c_{\pm}| < |c^{ps}_{\pm}| = \sqrt{a^2 - 1}$) the wider is the plateau.
When the plateau covers all
the range of variation of $\theta$ the states
become absolutely separable.
The case $|c_{\pm}| > |c^{ps}_{\pm}|$
would be the only one for which there is no point
of zero entanglement,
but it has to be discarded since the corresponding matrices do
not fulfill
the Heisenberg principle.
For simmetric mixed states such that
$V \neq V_s$ the entanglement calculations
give results that are similar
to the ones obtained for the  pure states case (see Fig.\ref{mns}): unitary redefinitions of the modes
can increase the entanglement. In addition, as described above, $E_{\cal{N}}$ can vanish
in correspondence of a whole interval of mode redefinitions.

\begin{figure}[h]
\includegraphics[height=5cm, width=8cm, clip]{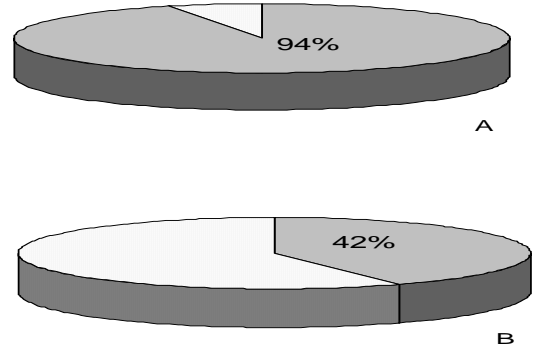}
\caption{ Statistical results for gaussian mixed states with
covariance matrix in standard form (A) and $V \neq V_s$ (B). The
parameters range is the following: $V_{ij} \in [-1,1]$ for $i \neq
j$ while $V_{ii} \in [0.5,1.5]$. The rate of absolutely separable in
first case (A) states is about 94\% over $7746$ generated states,
while in the second case (B) the rate of absolutely separable states
is lower (about $42\%$ over $4692$ states). }
 \label{stat1}
\end{figure}

Let us now turn to analyze the case of non-symmetric mixed states,
starting as usual from those whose covariance matrices are in standard form. A great
number of mixed states, in the defined range, turn out to be
absolutely separable i.e., $E^{+}(\rho)=0$.
Fig.\ref{stat1}A shows
the amount of absolutely separable states with respect to the amount
of randomly generated physical states. Although the chosen range is
not representative of the whole mixed gaussian states space i.e.,
different choices would  have led to different results, it  allows
us to make some general remarks.

The high separability rate can be understood by using the Simon
criterion~\cite{Sim}: it is sufficient for $c_{\pm}$ to have the
same sign to achieve separability. This set of absolutely separable
mixed states includes all the classical states defined in
Ref.~\cite{X-b,Oliv}. On the other hand, when the mixed state are
not absolutely separable there always exists,  as in the pure states
case, a non-void set of unitary modes redefinitions for which the
entanglement is zero. This result, together with the previously
obtained for pure states, allows one to claim that {\em no
absolutely entangled bipartite gaussian states exist}. Every
gaussian state can be regarded as separable, it is just a matter of
choosing the right operators/modes redefinition. The maximum
entanglement is reached by pure states, while mixed ones have, in
general a low logarithmic negativity.

We finally consider the case of non-symmetric  mixed states for
which $V \neq V_s$. The new statistics (Fig.\ref{stat1}B), within
the same parameters range, shows that, in comparison with the $V =
V_s$ case, the set of states that have nonzero entanglement for some
bipartition is far larger. This is due, as in the pure states case,
to the fact that entanglement can be increased with a proper choice
of mode redefinition and it is true even for initially separable
states. We remind that a different choice of the range would lead in
general to different percentage rates.

\section{Conclusions}
\label{sec:conclu}
The main purpose of this paper
is to show how the conceptual issue of entanglement relativity
i.e., the crucial dependence of entanglement on the choice of subsystems, can be addressed in a quantitative
fashion for a specific,
yet quite important, class of quantum states. We have studied how the entanglement of a
two-modes bosonic gaussian state changes when
the modes are redefined through  number-conserving unitary
transformations.
This redefinitions have to be seen analogously to "passive" rotations in elementary geometry:
the quantum state is unaffected, it is just the way it is split into subsystems i.e., modes, that
changes.
One  obtains a set of  bi-partitions parametrized by $2$ by $2$
unitary matrices; this makes the choice of the set of possible bipartitions, typically a huge
and rather ill-defined one,
easy to be dealt with analytical and numerical techniques.

We have found that
with a suitable unitary redefinition of the modes every
gaussian state can be regarded as separable. This means that {\em
there are no absolutely entangled bipartite gaussian states} i.e.,
for the chosen class of tensor product structures the function
$E^{-}(\rho)$ (see Eq. \ref{abs-meas}) is identically vanishing.
Moreover a large class of mixed gaussian states can be considered as
absolutely separable, $E^{+}(\rho) \equiv 0$, meaning that no
unitary  redefinition of the modes can 'entangle' these states.

This kind of entanglement variation analysis can be extended to a larger set of Bogolioubov
transformations: those that do not conserve the particle number and that are
able to induce further bipartitions \cite{nota1}.

% ***
%Moreover the set of unitary transformations can be extended to the
%generic commutation relations conserving ones: $a_i \rightarrow U a_i U^{-1}$. We have just
%performed the so-called 'passive' transformations, which in general do not involve
%any kind of evolution (unitary or not) of the state: $\rho' = U\rho U^{-1}$.
Of course even upon this extension the main result about  the
non-existence of absolutely entangled gaussian states will remain
valid, whereas the set of absolutely separable states might become
smaller.  Finally, recent papers~\cite{fermion1,fermion2} have
addressed the problem of dealing with gaussian states even in
fermionic systems.  This naturally suggests the possibility of
extending our investigations to the fermionic realm.

\section{acknowledgments.}
We are grateful to Zhao Yang, F. Illuminati, M.G.A. Paris and F.
Rossi for fruitful discussions and suggestions.

\newpage


\begin{thebibliography}{xx}
\bibitem{nota}
For this equation only, following \cite{symple}, we changed the index order in the definition of $V$
by setting $x_1 = q_A, x_2 = q_B, x_3 = p_A, x_4 = p_B$; in this way $S_U$ can be compactly defined
in terms of U.
\bibitem{nota1} We would like to stress these number non-conserving transformations are conceptually different
from the ones formerly  used in this paper ($T$s). In fact the latter are merely used in the algorithm to generate new
covariance matrices  from each given $V_s$.
\end{thebibliography}
\end{document}